# MicroRNAs Implicated in Dysregulation of Gene Expression Following Human Lung Transplantation

Wei Zhang[1,2], Tong Zhou[3,4], Shwu-Fan Ma[5], Robert F. Machado[3,4], Sangeeta M. Bhorade[5]*, Joe G.N. Garcia[3,4]*

[1]Department of Pediatrics; [2]Institute of Human Genetics; [3]Institute for Personalized Respiratory Medicine; [4]Section of Pulmonary, Critical Care & Sleep Medicine, Department of Medicine, University of Illinois, Chicago, IL 60612; [5]Section of Pulmonary/Critical Care, Department of Medicine, University of Chicago, Chicago, IL 60637

*Correspondence and requests for reprints should be addressed to:

Joe G.N. Garcia MD, Earl M Bane Professor of Medicine, Pharmacology and Bioengineering, University of Illinois Hospital and Health Sciences System, 914 South Wood St. (MCA101). Chicago, IL 60612; Tel: 312-996-9450; Fax: 312-413-0238; E-mail: jggarcia@uic.edu

Sangeeta Bhorade, MD, Associate Professor of Medicine, Medical Director, Lung Transplantation Program, University of Chicago, 5841 S. Maryland Avenue, MC 6076, Chicago, IL 60637; Tel: 773-834-1119; Fax: 773-702-6500; E-mail: sbhorade@medicine.bsd.uchicago.edu

**Running Title**: Gene dysregulation in lung transplants

**Abstract**

**Background**: Lung transplantation remains the only viable treatment option for the majority of patients with advanced lung diseases. However, 5-year post-transplant survival rates remain low primarily secondary to chronic rejection. Novel insights from global gene expression profiles may provide molecular phenotypes and therapeutic targets to improve outcomes after lung transplantation. **Methods:** Whole-genome gene expression profiling was performed in a cohort of patients that underwent lung transplantation as well as healthy controls using the Affymetrix Human Exon 1.0ST Array. To explore the potential roles of microRNAs (miRNAs) in regulating lung transplantation-associated gene dysregulation, miRNA expression levels were also profiled in the same samples using the Exiqon miRCURY™ LNA Array. **Results:** In a cohort of 18 lung transplant patients, 364 dysregulated genes were identified in Caucasian lung transplant patients relative to normal individuals. Pathway enrichment analysis of the dysregulated genes pointed to Gene Ontology biological processes such as "defense response", "immune response" and "response to wounding". We then compared the expression profiles of potential regulating miRNAs, suggesting that dysregulation of a number of lung transplantation-associated genes (e.g., *ATR*, *FUT8*, *LRRC8B*, *NFKBIA*) may be attributed to the dysregulation of their respective regulating miRNAs. **Conclusions:** Following human lung transplantation, a substantial proportion of genes, particularly those genes involved in certain biological processes like immune response, were dysregulated in patients relative to their healthy counterparts. This exploratory analysis of the relationships between miRNAs and their gene targets in the context of lung transplantation warrants further investigation and may serve as novel therapeutic targets in lung transplant complications.

**Background**

For many patients with end-stage lung diseases, lung transplantation is often the only remaining viable therapeutic measure (1) . The number of lung transplants is ~1,500 on average a year in the United States, which represents ~45% of lung transplants performed world-wide (2). Compelling data have documented the beneficial impact of lung transplantation on functional status, hemodynamics, and quality of life. Less compelling, however, is the demonstration of a survival benefit due to significant constraints on long-term survival (1).

Although short-term survival has improved via improved surgical techniques, donor preservation and immunosuppressive agents, long-term survival remains reduced after lung transplantation. The major cause of decreased long-term survival is bronchiolitis obliterans syndrome (BOS), a physiological measure of chronic rejection after lung transplantation. Approximately 50% of lung transplant recipients will develop BOS by five years post transplantation. However, the pathogenesis of BOS has not been clearly elucidated. Both alloimmune-dependent and -independent factors have been suggested to contribute to BOS pathogenesis. These factors include acute rejection, lymphocytic bronchiolitis, acute infectious etiologies and gastroesophageal reflux disease.

In addition, an individual patient's genetic make-up may also contribute to the prognosis after lung transplantation as well as to the development of various complications. Particularly, global gene expression profiling has been used to identify unique expression signatures in organ transplant biopsies that may help distinguish various outcomes such as acute rejection, acute dysfunction without rejection and well-functioning transplants with no rejection history (3-5). For example, gene expression in bronchoalveolar lavage cell samples from lung transplant recipients with and without acute rejection on simultaneous lung biopsies was examined and specific expression patterns were demonstrated at defined time points after transplantation in allografts (4). Though not definitive and comprehensive, these studies showed the potential power of whole-genome microarrays to identify biomarkers of acute/chronic transplant rejection and development of other complications.

Notably, gene expression itself has been demonstrated to be a complex and quantitative trait that varies within and between natural human populations (6-11) and is controlled by various genetic, epigenetic and non-genetic factors (12-15). MicroRNAs (miRNAs), small (21-25nt) non-coding RNA molecules, have emerged as a novel class of gene regulators that may affect various complex phenotypes including disease susceptibility and drug response (16, 17). Integrating whole-genome mRNA and miRNA profiles, therefore, could help elucidate the complex cellular response and its mechanisms in lung transplant patients, and provide novel biomarkers for the outcomes of lung transplantation. Specifically, we compared whole-genome transcriptional expression data profiled using the Affymetrix Human Exon 1.0ST Array (exon array) in peripheral blood mononuclear cells (PBMCs) from lung transplant patients and normal individuals. We searched for any enriched pathways or biological processes involved the dysregulated genes in lung transplant patients. We further demonstrated that miRNAs could potentially play a critical role in determining the gene expression dysregulation observed in lung transplant patients.

**Methods**

*Subjects and collection of PBMC samples*

The study was approved by the Institutional Review Board of each collaborating institution with written informed consent obtained from all subjects, and was performed in accordance with the principles in the Declaration of Helsinki. PBMC samples were collected from 18 patients (14 Caucasian Americans and 4 African Americans) who underwent single-lung transplant or bilateral single-lung transplant surgery during 2005-2008. The PBMC samples were collected 1.25-29.75 months after lung transplantation, representing a general patient population at the sample collection time. Control samples were collected from healthy individuals (27 Caucasians and 8 African Americans) with no diagnosis of pulmonary disorders at The University of Chicago Medical Center. **Table 1** shows the clinical characteristics of the study cohort.

*Obtaining exon array data*

Total RNA was extracted from PBMCs and prepared using standard molecular biology protocols. RNA concentration and purity was determined and before gene expression profiling using the Affymetrix Human Exon 1.0ST Array (exon array) (Affymetrix, Inc., Santa Clara, CA). The microarray labeling, hybridization and processing was performed at the University of Chicago Microarray Core Facility according to the manufacturer's protocol.

*Processing of exon array data*

We used the experimental probe masking workflow provided by the Affymetrix Power Tools v.1.12.0 (http://www.affymetrix.com/) to filter the probeset (exon-level) intensity files by removing probes that contain known SNPs in the dbSNP database v129 (18, 19). The resulting probe signal intensities were quantile normalized over all samples, summarized with the robust multi-array average (RMA) algorithm (20) and $\log_2$ transformed with a median polish (19) for ~22,000 transcript clusters (gene-level) with the core set (i.e., with RefSeq-supported annotations) (21). Adjustment for possible batch effect was conducted by COMBAT (http://jlab.bu.edu/ComBat/) (22). We consider a transcript cluster to be reliably expressed if the DABG (detection above ground) (23) p-value computed by the Affymetrix Power Tools was less than 0.01 in at least 80% of the samples in each test group (healthy controls or patients) in each population, respectively. We further limited our analysis set to the genes with unambiguous annotations by Affymetrix. Totally, 11,461 and 11,576 transcript clusters in the Caucasian American and African American samples, respectively, met these criteria and were further analyzed. We have deposited the raw and processed exon array data in the NCBI Gene Expression Omnibus (GEO) (Accession Number: GSE49081).

*Obtaining miRNA expression data*

The expression levels of miRNAs were profiled using the Exiqon miRCURY™ LNA Array v10.0 (~700 human miRNAs, updated to miRBase 11.0 annotation) (24) (Exiqon, Inc., Denmark). Briefly, total RNA from PBMCs was extracted and prepared according to manufacturer's protocol. Array hybridization was performed by Exiqon with the quantified signals background corrected using *normexp* with offset value 10 based on a convolution model (25) and normalized using the global Lowess regression algorithm. In total, 318 miRNAs and 309 miRNAs were found to be expressed in the Caucasian American samples and the African American samples, respectively (i.e., present in at least 80% of total samples in each population).

*Identifying genes dysregulated in patients with lung transplants*

We excluded genes on chromosomes X and Y to avoid the potential confounding effect of gender. SAM (Significance Analysis of Microarrays) (26), implemented in the *samr* library of the R Statistical Package (27), was used to identify differential genes between patients who underwent lung transplantation and healthy controls in the Caucasian American and African American samples, respectively, as well as between patients with and without development of BOS. Transcripts with a greater than 1.5 fold-change and q-value (28) less than 0.01 (i.e., 1% FDR, false discovery rate) were deemed significantly dysregulated. We searched for any enriched pathways and biological processes among the differential genes relative to the respective analysis set using the DAVID (Database for Annotation, Visualization and Integrated Discovery) tool (29, 30). The following databases were included: KEGG (Kyoto Encyclopedia of Genes and Genomes) (31), BioCarta (http://www.biocarta.com/), Reactome (32), PANTHER (33), and Gene Ontology (GO) (34). Due to the exploratory nature of this study, we chose to use a relatively lenient cutoff, i.e., FDR<25% after the Benjamini-Horchberg procedure

(35) and a minimum of 5 differential genes in a pathway or biological process, for the DAVID analysis.

*Identifying relationships between dysregulated genes and potential regulating miRNAs*

The differential genes were then searched against the MicroCosm Targets (24) (i.e., miRanda algorithm) through our ExprTarget database (http://www.scandb.org/apps/microrna/) (36) for potential regulating miRNAs (miRanda $p<1.0\times10^{-4}$). Only human miRNAs that are expressed in these samples (318 miRNAs in the Caucasian samples; 309 miRNAs in the African American samples) were included in the analysis. The expression patterns of those miRNAs and their corresponding gene targets were compared between patients and normal controls using standard *t*-test. Significant miRNA-mRNA relationships (i.e., negative association between miRNA and mRNA at *t*-test $p<0.05$) were further confirmed using linear regression. The Pearson correlation coefficients and the associated p-values (cutoff $p<0.05$) were calculated using the *lm* library of the R Statistical Package (27).

**Results**

*Identifying genes dysregulated in patients with lung transplants*

In total, 364 genes were differentially expressed between 14 Caucasian American patients with lung transplants (n=14) and normal controls (n=27) (fold-change>1.5, q-value<0.01) with 292 down-regulated and 72 up-regulated genes (**Figure 1**) (**Supplementary Table S1**). By comparison, only four genes were dysregulated between African American cases (n=4) and normal black controls (n=8) (fold-change>1.5, q-value<0.01) with one down-regulated gene and three up-regulated genes. In addition, no significant BOS-associated genes were identified at the q-value<0.01 level between Caucasian patients with (n=5) and without BOS (n=9), probably limited by the small sample size of this group of patients in our study cohort. At a looser cutoff (q-value<0.05), this comparison revealed 884 up-regulated genes associated with BOS, indicating a potential trend of substantial dysregulation in patients who developed BOS. A larger sample size may be needed for a more reliable description of this comparison. We focused our downstream analyses on the more robust list of dysregulated genes in all patients with lung transplants relative to healthy controls, particularly in the Caucasian samples.

*Enriched pathways among dysregulated genes*

DAVID analysis on the 364 dysregulated genes in Caucasian American cases revealed 12 enriched pathways and GO biological processes such as "response to bacterium", "immune response" and "response to wounding" (FDR<25%, a minimum of 5 genes). Six known pathways and GO biological processes such as "hemostasis" and "blood coagulation" were enriched among the 292 down-regulated genes, while 19 GO biological processes such as "defense response", "response to bacterium", and "immune response" were enriched among the 72 up-regulated genes (**Supplementary Table S2**). **Table 2** shows the top-ranking pathways and biological processes (FDR<10%, a minimum of 10 genes) for each gene group (i.e., "all dysregulated", "down-regulated" and "up-regulated"). In addition, analysis on the 884 potentially BOS-associated genes showed enrichment in pathways such as "T-cell receptor signaling pathway" and "apoptosis".

*Identifying potential regulating miRNAs for the dysregulated genes*

We searched for potential regulating miRNAs for the dysregulated genes in lung transplant patients based on the predictions of the miRanda algorithm (24). Among the 292 down-regulated genes in the Caucasian American patients, 178 miRNA-mRNA relationships corresponding to 95 expressed miRNAs and 78 genes were identified, while 74 miRNA-mRNA relationships corresponding to 40 expressed miRNAs and 31 genes were identified in the 72-up-regulated genes (miRanda $p<1.0\times10^{-4}$). In comparison, nine miRNAs were identified for the single down-regulated gene, *SMOX* (encoding spermine oxidase) in the African American patients (miRanda $p<1.0\times10^{-4}$).

We further searched for miRNAs that showed a negatively associated expression pattern with their potential gene targets using *t*-test (**Table 3**). For the down-regulated genes in the Caucasian American patients, six potential regulating miRNAs were found to be up-regulated in the patients (corresponding to 8 miRNA-mRNA relationships). For example, hsa-miR-34a was up-regulated (*t*-test p=0.0001), while its gene target, *FUT8* (encoding fucosyltransferase 8) was down-regulated in transplant patients; and hsa-miR-519e was up-regulated (*t*-test p=0.003), consistent with down-regulation of its potential targets, *ATR* (encoding ataxia telangiectasia and Rad3 related) and *PYHIN1* (encoding pyrin and HIN domain family, member 1). In contrast, six miRNA-mRNA relationships were identified among the up-regulated genes in the patients. For example, hsa-miR-381 was down-regulated while its potential target gene *NFKBIA* (encoding nuclear factor of kappa light polypeptide gene enhancer in B-cells inhibitor, alpha), was up-regulated in the patients. (*t*-test p=0.00067). Linear regression confirmed the relationships between miRNAs and their potential gene targets (**Table 3**). Figure 2 shows some examples of the confirmed (p<0.05) miRNA-mRNA relationships in the Caucasian American samples.

**Discussion**

Lung transplantation is associated with major complications such as infection, acute rejection and chronic rejection characterized by BOS (2). Elucidating the complex cellular and physiological response after lung transplantation will be critical to understanding the pathogenesis of acute and chronic complications after lung transplantation. To our knowledge, this is the first study to assess the relationship between dysregulated genes and potential gene regulators of miRNAs in patients that underwent lung transplantation.

Approximately 3% of the analyzed genes were differentially expressed in Caucasian patients with lung transplants, indicating systematical dysregulation of certain genes in these patients, potentially implicating their outcomes after lung transplantation. Using DAVID (29, 30), these lung transplant-associated genes were found to be enriched in a number of known pathways and GO (34) biological processes including "immune response", "defense response", "response to wounding", "hemostasis" and "blood coagulation" (**Table 2**). Interestingly, biological processes such as "blood coagulation" and "hemostasis" were enriched among down-regulated genes, while biological processes such as "immune response", "defense response", "response to bacterium", and "response to wounding" were enriched among up-regulated genes. Lung transplant patients are routinely anticoagulated to prevent thrombosis, and given antibiotic prophylaxis to prevent infections and immunosuppressants to prevent organ rejection. It appears that after lung transplantation and relevant treatments, genes related to blood coagulation were significantly down-regulated in patients, while genes related to the aftermath of a major surgery including "response to wounding" were significantly up-regulated in patients.

Notably, many of these pathways shared a significant number of genes, displaying the complex interactions of several biological processes after lung transplantation. For example, all of the seven up-regulated genes including *FOS* (encoding FBJ murine osteosarcoma viral oncogene homolog), *IL8* (encoding interleukin 8), *IL1B* (encoding interleukin 1, beta), involved in "inflammatory response" were also involved in "response to wounding". In addition, many genes related to the response to bacterial infection and lipopolysaccharide (LPS) were up-regulated in patients such as *NFKBIA*, *FOS*, *PTGS2* (encoding prostaglandin-endoperoxide synthase 2), *ADM* (encoding adrenomedullin), *SOCS3* (encoding suppressor of cytokine signaling 3), *TRIB1* (encoding tribbles homolog 1) and *IL1B*. Among them, *FOS*, *PTGS2* and *IL1B* are genes involved in "response to glucocorticoid stimulus", which was also enriched in the up-regulated genes and likely reflected treatment. Besides these up-regulated immune response genes, 20 other immune response genes such as *CCR4* (encoding chemokine receptor 4), *CD86* (encoding CD86 molecule), were down-regulated in patients. Since lung transplant patients were treated continuously with immunosuppressive drugs, the dysregulation of some of these immune response genes could be due to the on-going immunosuppressant treatments. Obviously, some immune response genes such as *FOS*, *PTGS2*, *IL1B* could be induced by immunosuppressive drugs (e.g., glucocorticoids), while some other immune response genes could be suppressed by drug treatments. Dysregulation of these genes after transplantation provides more insight regarding the interactions of various biological processes and may ultimately provide biomarkers of the various complications related to outcomes after transplantation.

Among the 364 differential genes in patients, a number of genes showed an expression pattern correlated with their potential regulating miRNAs (**Table 3**). Using a linear regression model, we demonstrated that expression of specific miRNAs was significantly correlated with the expression levels of their potential gene targets. For example, *ATR* and *LRRC8B* (encoding leucine rich repeat containing 8 family, member B) were down-regulated in transplant patients. Their expression levels were significantly correlated with their potential regulating miRNA, respectively (**Figure 2**). The majority of the identified miRNA-mRNA relations could be confirmed using linear regression. Notably, the gene *NFKBIA* was found to be negatively associated with its potential regulating miRNA has-miR-381 (**Table 3**). *NFKBIA* is also involved in "response to wounding" and "response to LPS", suggesting that miRNAs may contribute to these biological processes in lung transplant patients. Our results suggest that the complex dynamics of dysregulated genes in these patients may be partially attributed to the differential expression of their potential regulating miRNAs following lung transplantation, as well as relevant treatments such as immunosuppressive drugs, anticoagulants.

We recognize that there are some limitations to this exploratory study. Firstly, the sample size of the population was small and therefore, this analysis must be validated with a larger cohort of patients. Secondly, some potential confounding factors (e.g., types of immunosuppressive agent that the patients are taking) might influence gene dysregulation. In addition, we were unable to compare our findings between different ethnic populations, as well as derive more robust conclusions for BOS-associated dysregulation, given the small sample size. However, given the exploratory nature of this analysis, our primary goal was to determine the putative relationships between dysregulated genes and regulating microRNAs in lung transplantation. Indeed, we were able to show a significant number of miRNA-mRNA relationships, suggesting that the regulation of gene targets by these miRNAs in the context of lung transplantation warrant further investigation, and could ultimately provide as novel therapeutic targets in lung transplant complications.

**Acknowledgements**

This work was supported by an NIH grant, HL058064. The funding body had no role in study design, data collection and analysis, decision to publish, or preparation of the manuscript

**Competing interests**

None of the authors has a financial relationship with a commercial entity that has an interest in the subject of the presented manuscript or other conflicts of interests.

**Authors' contributions**

Conception and design: WZ, RFM, SB, JGNG. Analysis and interpretation: WZ, TZ, RFM, SB, JGNG. Sample preparation and experiments: SM. Drafting the manuscript for important intellectual content: WZ, TZ, RFM, SB, JGNG. All authors read and approved the final manuscript.

**Table 1.** Summary of the study cohort.

|  | Caucasian | Caucasian controls | African American | African American controls |
|---|---|---|---|---|
| **Age (mean ± SD) years** | 56 ± 18 | 65 ± 12 | 48 ± 9 | 75 ± 7 |
| **Gender (F) (%)** | 3 (21%) | 6 (22%) | 2 (50%) | 5 (63%) |
| **Underlying Diagnosis** | | -- | | -- |
| *COPD* | 4 | | 1 | |
| *IPF* | 7 | | 1 | |
| *CF* | 1 | | 0 | |
| *Bronchiectasis* | 1 | | 0 | |
| *Sarcoidosis* | | | 1 | |
| *PLCH* | 1 | | 0 | |
| *CTD-ILD* | | | 1 | |
| **Type of transplant** | | -- | | -- |
| *SLT* | 8 | | 1 | |
| *BLT* | 6 | | 3 | |
| **Time from transplant (mean ± SD) months** | 10.4 ± 9.1 | -- | 11 ± 4 | -- |
| **BOS** | 5 (36%) | -- | 1 (25%) | -- |

F, female; COPD, chronic obstructive pulmonary disease; IPF, idiopathic pulmonary fibrosis; SLT, single-lung transplant; BSL, bilateral single-lung transplant; BOS, bronchiolitis obliterans syndrome; PLCH, Pulmonary Langerhans Cell Histiocytosis; CTD, connective tissue disease; ITD, interstitial lung disease

**Table 2**. Some enriched pathways and biological processes among the dysregulated genes in Caucasian American patients.

| Gene Group | Pathway Category | Count | P | Fold Enrichment | FDR |
|---|---|---|---|---|---|
| *all* | GO:0009617~response to bacterium | 16 | 1.09E-06 | 4.74 | 0.002 |
| | GO:0006955~immune response | 33 | 1.39E-06 | 2.56 | 0.001 |
| | REACT_604:Hemostasis | 20 | 3.31E-05 | 2.81 | 0.001 |
| | GO:0009611~response to wounding | 26 | 8.98E-05 | 2.37 | 0.054 |
| | GO:0006952~defense response | 26 | 1.40E-04 | 2.31 | 0.063 |
| *down-regulated* | REACT_604:Hemostasis | 18 | 4.50E-05 | 2.96 | 0.002 |
| | GO:0050817~coagulation | 10 | 6.18E-05 | 5.61 | 0.083 |
| | GO:0007596~blood coagulation | 10 | 6.18E-05 | 5.61 | 0.083 |
| | GO:0007599~hemostasis | 10 | 7.70E-05 | 5.46 | 0.052 |
| *up-regulated* | GO:0009617~response to bacterium | 10 | 4.11E-08 | 13.25 | 0.000 |
| | GO:0006955~immune response | 13 | 1.77E-05 | 4.52 | 0.004 |
| | GO:0006952~defense response | 12 | 2.76E-05 | 4.76 | 0.005 |
| | GO:0007610~behavior | 10 | 8.23E-05 | 5.30 | 0.009 |
| | GO:0010033~response to organic substance | 12 | 5.42E-04 | 3.41 | 0.040 |
| | GO:0009611~response to wounding | 10 | 5.80E-04 | 4.08 | 0.040 |

GO, Gene Ontology biological process; REACT, Reactome pathway; FDR, false discovery rate

**Table 3**. Some dysregulated genes in the Caucasian American patients are negatively associated with potential regulating miRNAs.

| Gene Class | Target Gene | Gene Title | microRNA | p (miRanda)[a] | p (t-test)[b] | p (linear regression)[c] | $r^2$ (linear regression)[d] |
|---|---|---|---|---|---|---|---|
| *down-regulated* | *TCF4* | transcription factor 4 | hsa-miR-299-3p | 0.000031 | 0.012 | 0.035 | 0.11 |
| | *LRRC8B* | leucine rich repeat containing 8 family, member B | hsa-miR-29b-1* | 0.000056 | 0.019 | 0.0057 | 0.20 |
| | *C14orf2* | chromosome 14 open reading frame 2 | hsa-miR-34a | 0.000058 | 0.0001 | 0.042 | 0.11 |
| | *FUT8* | fucosyltransferase 8 | hsa-miR-34a | 0.0000003 | 0.0001 | 0.0084 | 0.17 |
| | *C14orf135* | chromosome 14 open reading frame 135 | hsa-miR-451 | 0.0000086 | 0.0064 | 0.011 | 0.16 |
| | *ATR* | ataxia telangiectasia and Rad3 related | hsa-miR-519e | 0.0000075 | 0.0031 | 0.0046 | 0.19 |
| | *PYHIN1* | pyrin and HIN domain family, member 1 | hsa-miR-519e | 0.000091 | 0.0031 | 0.0074 | 0.17 |
| | *TCF4* | transcription factor 4 | hsa-miR-629 | 0.0000071 | 0.022 | 0.045 | 0.12 |
| *up-regulated* | *CA1* | carbonic anhydrase I | hsa-miR-590-5p | 0.000052 | 0.00067 | NS | |
| | *NFKBIA* | nuclear factor of kappa light polypeptide gene enhancer in B-cells inhibitor, alpha | hsa-miR-381 | 0.000021 | 0.0012 | 0.023 | 0.13 |
| | *NFIL3* | nuclear factor, interleukin 3 regulated | hsa-miR-374a | 0.000046 | 0.011 | NS | |
| | *DOCK4* | dedicator of cytokinesis 4 | hsa-miR-28-5p | 0.000095 | 0.012 | NS | |
| | *PLK2* | polo-like kinase 2 | hsa-miR-126 | 0.000055 | 0.025 | NS | |
| | *PLK2* | polo-like kinase 2 | hsa-miR-27b | 0.000000088 | 0.033 | NS | |

a: p-values from the miRanda algorithm; b: p-values by comparing miRNA expression levels between patients and normal controls; c: p-values from linear regression tests on the relationships between miRNAs and target genes; NS, not significant; d: correlation coefficients from the linear regression tests

**Figure Legends**

**Figure 1**. Genes dyregulated in the Caucasian American lung transplant patients.

In total, 364 genes were differentially expressed between lung transplant patients and normal controls. Among them, 292 genes were down-regulated in the patients, while 72 genes were up-regulated in the patients. Blue represents down-regulation. Red represents up-regulation. Red bars: controls; Green bars: lung transplant patients.

**Figure 2**. Examples of the relationships between potential regulating miRNAs and the dysregulated genes.

The p-values were from linear regression tests. Blue dots represent the control samples. Red dots represent the patient samples. X-axis: miRNA expression; Y-axis: mRNA expression. (A). *ATR* (encoding ataxia telangiectasia and Rad3 related) was down-regulated in the patients; (B). *LRRC8B* (encoding leucine rich repeat containing 8 family, member B) was down-regulated in the patients; (C). *PYHIN1* (encoding pyrin and HIN domain family, member 1) as down-regulated in the patients; and (D). *NFKBIA* was up-regulated in the patients.

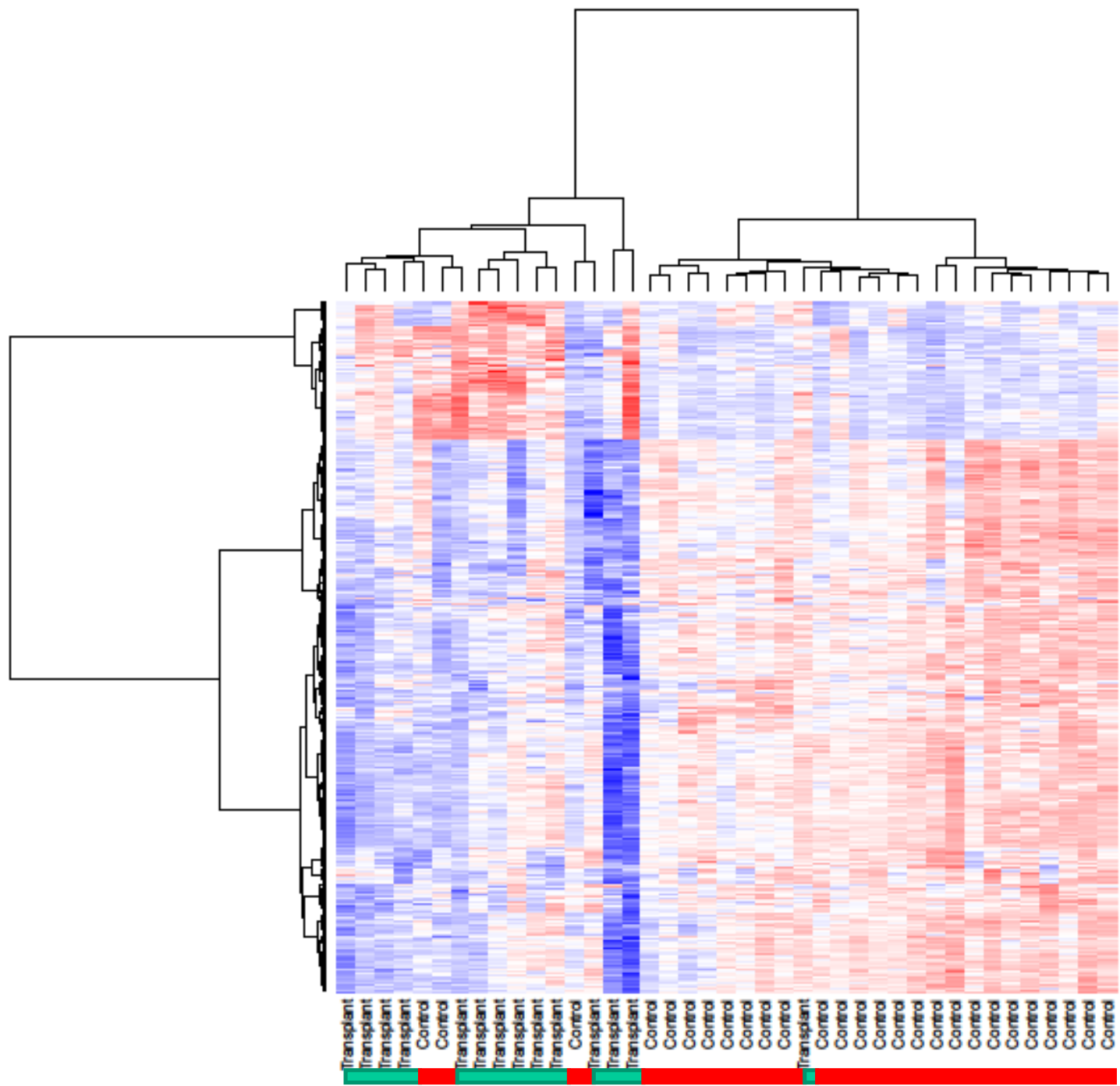

Fig. 1

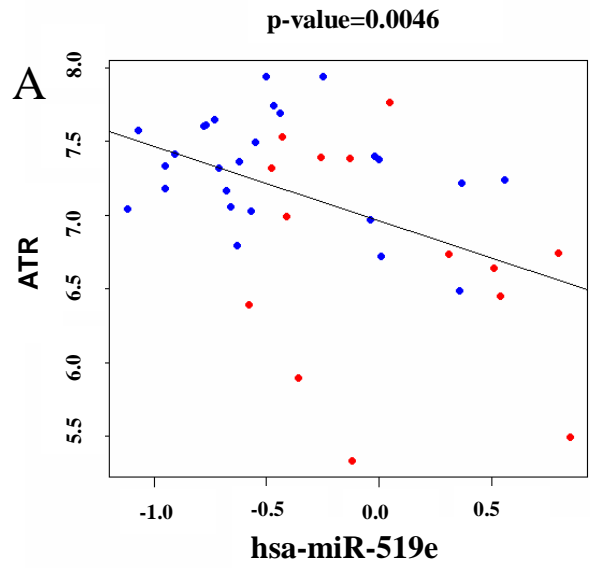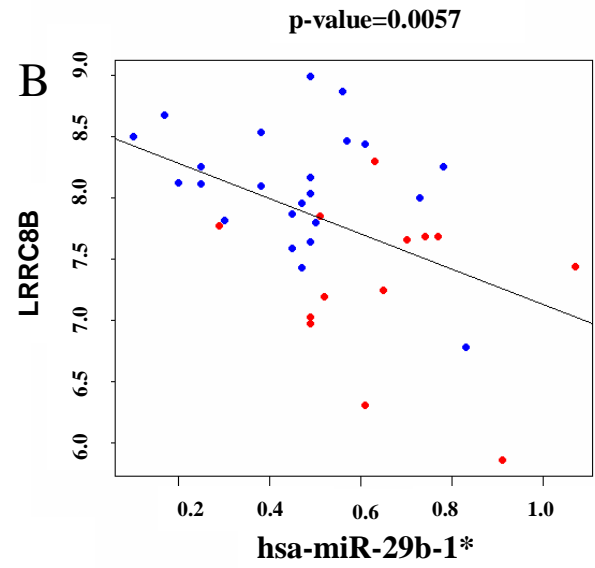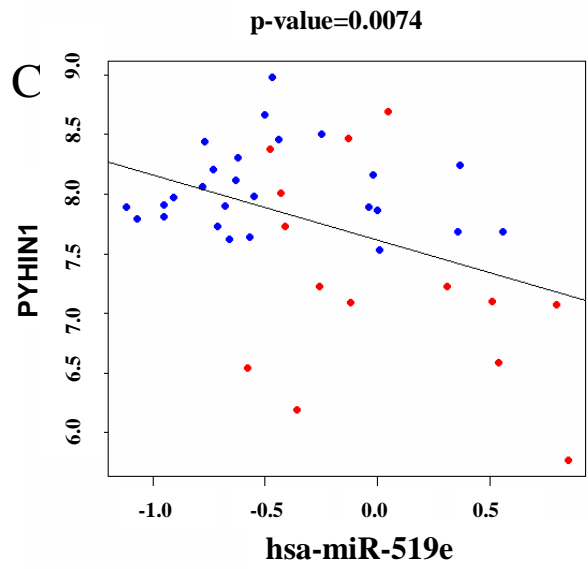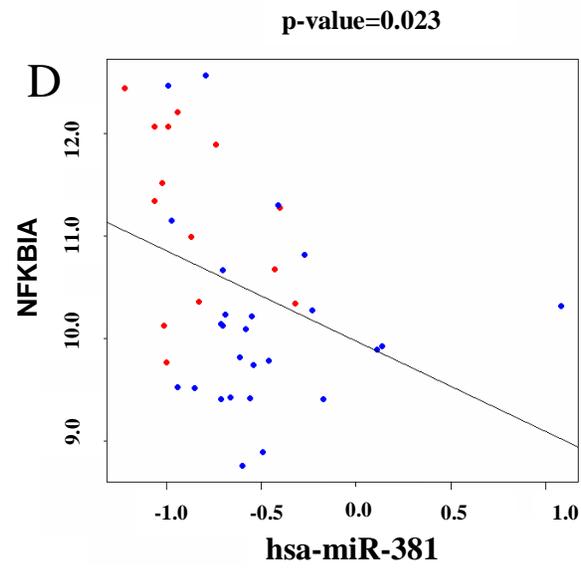

**Fig. 2**